\documentclass[10pt]{article}
\usepackage{graphicx}
\usepackage{times}
\usepackage{fullpage}
\usepackage{amssymb}
\usepackage{graphicx}
\usepackage{latexsym}
\usepackage{threeparttable}
\usepackage{booktabs}
\usepackage{bbding}
\usepackage{epsfig}
\usepackage{algorithm}
\usepackage[noend]{algpseudocode}
\usepackage{makecell}
\usepackage{colortbl}
\usepackage{rotating}
\usepackage{tabularx}
\usepackage{graphicx}
\usepackage{amsmath}
\usepackage{makeidx}  %
\usepackage{color}    %
\usepackage{epsfig}
\usepackage{multirow}
\usepackage{enumerate}
\usepackage{threeparttable}
\usepackage[small,bf]{caption}
\usepackage{color}
\definecolor{darkgreen}{RGB}{47,109,79}
\definecolor{darkblue}{RGB}{57,79,99}

\usepackage[colorlinks=true, linkcolor = darkgreen, citecolor = darkgreen, urlcolor = darkblue, filecolor = blue]{hyperref}

\makeatletter
\def\BState{\State\hskip-\ALG@thistlm}
\makeatother

\newcommand{\descr}[1]{\medskip\noindent\textbf{#1}}

\begin{document}

\title{\bf Privacy-Preserving Genetic Relatedness Test\thanks{A preliminary version of this paper appears in the Proceedings of the 3rd International Workshop on Genome Privacy and Security (GenoPri'16)}}

\author{Emiliano De Cristofaro$^1$, Kaitai Liang$^2$, Yuruo Zhang$^1$\\[1ex]
{\normalsize $^1$ Department of Computer Science, University College London, UK}\\
{\normalsize $^2$ Department of Computer Science, Aalto University, Finland}}
\date{}
\maketitle

\begin{abstract}

An increasing number of individuals are turning to Direct-To-Consumer (DTC) genetic testing to learn about their predisposition to diseases, traits, and/or ancestry. DTC companies like 23andme and Ancestry.com have started to offer popular and affordable ancestry and genealogy tests, with services allowing users to find unknown relatives and long-distant cousins. Naturally, access and possible dissemination of genetic data prompts serious privacy concerns, thus motivating the need to design efficient primitives supporting private genetic tests. In this paper, we present an effective protocol for privacy-preserving genetic relatedness test (PPGRT), enabling a cloud server to run relatedness tests on input an encrypted genetic database and a test facility's encrypted genetic sample. We reduce the test to a data matching problem and perform it, ``privately'', using searchable encryption. Finally, a performance evaluation of hamming distance based PP-GRT attests to the practicality of our proposals. 
\end{abstract}

\section{Introduction}
\label{sec:introduction}
Over the past few years, advances in genomics and genome sequencing are not only enabling progress in medicine and healthcare, but are also bringing genetic testing to the masses,
as an increasing number of ``Direct To Consumer'' (DTC) companies have entered,  and sometimes disrupted, the market.
For instance, 23andme.com offers a \$99/\textsterling 129 assessment of inherited traits, genealogy, and congenital risk factors, through genotyping of a saliva sample posted via mail, and Ancestry.com also offers low-cost genotyping-based DNA tests.

In this paper, we focus a popular test offered by many DTC companies, namely, Genetic Relatedness Test (GRT). This is used to identify whether or not a pair of individuals are closely related, genetically speaking. The standard approach to relative identification is to detect the identity-by-descent (IBD) segments between the individuals, and further identify the degree of relatedness via the amount of shared IBD segments~\cite{Pemberton10}. The quantity of shared IBD segments is then detected from the phased haplotypes (note a haplotype is a group of genes within an organism that was inherited together from a single parent) of (a pair of) individuals, which are the specific groups of genes that a progeny inherits from one parent and consists of a fixed number of single nucleotide polymorphisms (SNPs are the most common form of DNA variation occurring when a single nucleotide differs between members of the same species or paired chromosomes of an individual~\cite{stenson}).

Nevertheless, GRT services available today require individuals to send their genetic data (in plaintext) to possibly untrusted DTC companies.
Furthermore, collected genetic data is often impossible to anonymize~\cite{Homer08} and hard to protect from intentional or accidental leakage. Privacy risks from individual genetic exposure have been studied extensively~\cite{AydayCHT15}, thus motivating the need to design a GRT algorithms that can operate without accessing genetic data in the clear and violating individuals' privacy.  

We focus on genomic data that has already been phased into haplotypes and assume that the haplotypes of a pair of individuals with the same length are both interpreted by letters ``A, G, C, T'', so that the problem of detecting IBD shared segments %
is reduced to the identification of the shared positions of two equal-length strings. Dynamic programming can then be used to calculate shared positions, e.g., edit distance, hamming distance and longest common subsequence (LCS) algorithms.  

\subsection{Related Work} Overall, the research community has dedicated a lot of attention to genomic privacy, 
and proposed cryptographic techniques for privacy-preserving genetic 
testing~\cite{Ayday_Med_Tech_Report_2012,ayday_NDSS13,ayday_Healthtech13,Baldi_CCS_2011,Canim_2012,danezis2014fast,WPES12,de2013secure,Wang:2009:PGC:1653662.1653703}. 
Prior work on secure (two-party) computation could also be used to protect privacy in relatedness test protocol building on  traditional cryptographic primitives like homomorphic encryption (e.g.~\cite{DijkGHV10}), private set intersection (e.g.~\cite{Pinkas0Z14}), and garbled circuits based secure computation (e.g.~\cite{PinkasSSW09}). Some existing two-party secure distance computation protocols could also come to help here, for example, private set intersection~\cite{CristofaroT12}, privacy-preserving approximating edit distance~\cite{WangHZTWB15} based on garbled circuit and oblivious transfer, oblivious transfer based hamming distance system~\cite{BringerCP13}, or a homomorphic computation of edit distance~\cite{CheonKL15}.  
These two-party computation tools can be extended into a cloud-based context involving a trusted cloud server (with fully access to an online database consisting of a number of individuals' genetic data) and a relatedness test service (client) with its own genetic information. The client could use a additively homomorphic encryption scheme, e.g., Paillier cryptosystem~\cite{Paillier99}, to encrypt a haplotype and send it (along with the public key) to the server. The server would then encrypt each haplotype in its database using the same encryption algorithm, ``subtract'' client's encryption by the encrypted haplotype, and return the results to the client. The latter could then decrypt and identify the quantity of shared positions by checking the number of 0's decrypted. 

Two recent works~\cite{Hormozdiari14,He13} open a new perspective for privacy-friendly GRT by using fuzzy encryption technique. In these systems, each individual first compresses its haplotype into a 0/1 string, called private genome sketch, and then ``encrypts'' the sketch by using a random row of a given error correct code matrix. One may detect if user $A$ is relative by downloading $A$'s encrypted sketch, and next ``decrypting'' the sketch with its own private genome sketch. If the decryption closely leads to a row in the matrix, the haplotypes of both individuals are approximately matched.     
However, all the aforementioned computation approaches do not really scale well in practice, as they all suffer from an important limitation, i.e., the client is burdened with heavy computation and communication overhead, as it has to download all related ``encrypted'' results from the server and perform a huge numbers of decryption to identify the relatedness. How to design scalable privacy-preserving GRT that can scale on both server and client side constitutes the main motivation for our work.

\subsection{Roadmap} This paper presents a novel, efficient Privacy-Preserving Genetic Relatedness Test (PPGRT) protocol that relies on the cloud's computational power. Generally speaking, we allow the cloud to perform GRT by only given an encrypted genetic database and an encrypted personal haplotype, as opposed to the traditional context where the cloud is able to fully access the database. Before instantiating the protocol, we first discuss how PPGRT protocol can be generically constructed on top of a public key searchable encryption (e.g.~\cite{BCOP04}) and a symmetric key encryption cryptosystems. Next, we propose a concrete construction building on searchable encryption technique~\cite{WWP07}. Specifically, we encode the haplotypes stored on the cloud server as a bunch of search trapdoors, and the haplotype uploaded by a test issuer (i.e., the client) is interpreted as a set of search indices. An IBD shared segment detection is then reduced to a searchable matching problem, i.e., finding a match indicating a shared segment. 

Note that the method does not leak information to cloud server even if the server knows the length of shared segment as long as the secrecy of haplotypes is guaranteed. It is worth mentioning that the paper is the first to explore searchable encryption technique into relatedness test. We hope this work will inspire a new perspective for the future genomic privacy research.%

\section{Privacy-Preserving Genetic Relatedness Testing (PPGRT)}
\label{sec:construction}

\subsection{System Framework and Definition}

\begin{figure*}[t]
	\centering
\includegraphics[width=3.5in]{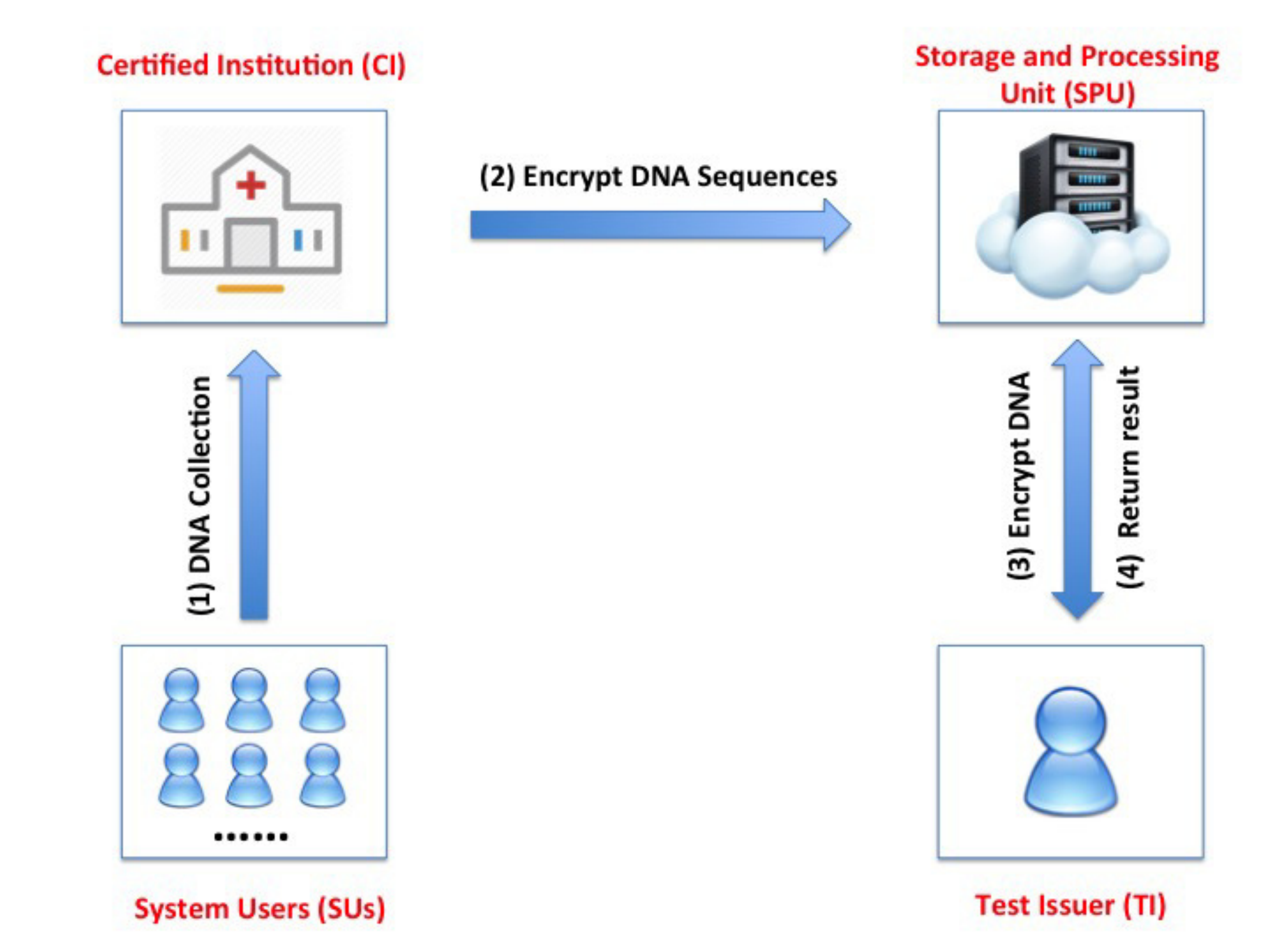}		
	\caption[System Structure]{System Model.}
	\label{fig:1}
\end{figure*}

We assume a genetic data sequence is represented as a haplotype consisting of five basic letters $\{A,G,C,T,*\}$, where $*$ denotes a piece of unknown/unmarked information, and assume the length of each letter is $l$. As illustrated in Fig.~\ref{fig:1}, there are four parties in a Genetic Relatedness Test (GRT) system, which mirrors the one introduced by Ayday et al. in~\cite{ARMFH13}, specifically:

\begin{enumerate}
\item {\bf System Users (SUs).} System users' plain genetic information are collected and phased into haplotypes by a trusted and certified institution, e.g. local health/medical center. 

\item {\bf Certified Institution (CI).} A trusted CI collects system users' plain genetic data to form a genome database ($GDB$), and encrypts $GDB$ to an $EDB$. Later, CI outsources $EDB$ to a storage and processing unit in which genetic relatedness testing is performed.   

\item {\bf Storage and Processing Unit (SPU).} A SPU stores $EDB$ locally and meanwhile, it is mainly responsible for running relatedness test as well as returning the test result to a test issuer. Note that SPU knows nothing about the underlying data stored in $EDB$.  

\item {\bf Test Issuer (TI).} A TI encrypts its phased haplotype via an \emph{asymmetric} key encryption mechanism, and sends the encryption to the SPU. After the test, the SPU returns the result to the TI without learning any (underlying haplotype) information.  
\end{enumerate}

\descr{Algorithms.} The system consists of the following algorithms:
\begin{enumerate}
\item $(sk, pk)\gets Setup(1^k)$: on input a security parameter $k$, output public parameter and secret key for the system.

 \item $EDB\gets GenEDB(sk, pk, GDB)$: on input secret key $sk$, public parameter $pk$ and a plain genome database, output an encrypted genome database $EDB$. We assume that $GDB$ consists of a list of plain haplotypes, i.e. $GDB=(h_i)_{i=1}^{d}$, and $EDB$ is a list of index/encrypted haplotype pairs, i.e. $EDB=(ind_i, H_i)_{i=1}^{d}$, where $|ind_i|=|h_i|$, and $d$ is the total number of haplotypes.
 
\item $C \gets GenQuery(pk, X)$: on input public parameter $pk$ and a plain haplotype information, output an encrypted haplotype $C$. 

\item $QL\gets Test(pk, EDB, C)$: on input public parameter $pk$, $EDB$ and an encrypted query, output a set $QL$ whereby the set includes $d$ test results between an encrypted query $C$ and each encrypted haplotype in $EDB$.        
\end{enumerate}

\subsection{Black-box PPGRT}
We highlight that Privacy-Preserving Genetic Relatedness Testing (PPGRT) can be generically constructed from a public key based searchable encryption (PKSE) and a symmetric encryption scheme. Using~\cite{BCOP04}'s definition for PKSE, and given a PKSE ($(SK, PK)\gets KeyGen(1^k)$, $S\gets PEKS(PK, W)$, $T\gets Trapdoor(SK, W)$ (note we assume that randomness is already taken in algorithms $PEKS$ and $Trapdoor$, so that each trapdoor and each index ``look'' uniformly random in the view of SPU.), $1/0\gets Test(PK, S, T)$), a symmetric encryption ($key\gets SSE.Key(1^k)$, $EDB\gets SSE.Enc(key, DB)$, $DB\gets SSE.Dec(key, EDB)$), we can build a generic PPGRT construction as follows: 
\begin{enumerate}
\item $Setup(1^k)$: run $KeyGen(1^k)$ algorithm to generate ($SK$, $PK$), and $SSE.Key(1^k)$ algorithm to generate $key$, and set $(SK, key)$ and $PK$ to be $sk$ and $pk$, respectively.  

 \item $GenEDB(sk, pk, GDB)$: run $SSE.Enc(key, h_i)$ algorithm $d$ times to encrypt each plaintext haplotype $h_i$ to become an encrypted value $H_i$, and further run $Trapdoor(SK, W_i^{j})$ algorithm to build a searchable trapdoor $T_i^j$ for each letter $W_i^j$ in a haplotype $h_i$, where $j\in [1,|h_i|]$. We have $EDB=(ind_i=(T_i^{1},....,T_i^{|h_i|})$, $H_i)_{i=1}^{d}$. 
 
\item $GenQuery(pk, X)$: run $PEKS(PK, W^{j})$ to generate a search index $S^j$ for each letter $W^{j}$ in the haplotype $X$, and further set $C=(S^1,...,S^{|X|})$, where $j\in [1, |X|]$. 

\item $Test(pk, EDB, C)$: given two-equal length ``sequences'' $C=(S^1, ..., S^{|X|})$ and $ind_i=(T_i^{1},....,T_i^{|h_i|})$, the calculation of shared (positions) length is reduced to the counting of the number of ``0'' output by the algorithm $Test$. Some dynamic distance programming algorithms (e.g. hamming/edit distance) can be used here. By intaking an index sequence $C$ and a trapdoor sequence $ind_i$ (with equal length), for $j=1$ to $|X|$, the algorithm checks the output of $Test(PK, S^j, T_i^{j})$. If the algorithm outputs 0, indicating there is a mismatch for the corresponding $j$-th position (of both sequences), we add 1 to the distance. We calculate the shared length (between $C$ and $H_i$) by subtracting $|X|$ with the distance, and store the result as an $i$-th tuple in $QL$. After $d$ rounds of the test, output $QL$.       
\end{enumerate}

\descr{PPGRT Security.} A secure PPGRT system needs to guarantee that: (1) $EDB$ must not leak the underlying encrypted sensitive haplotype data to SPU, (2) SPU cannot learn the underlying haplotype embedded in the search index $ind_i$, and (3) the underlying haplotype cannot be compromised from the test result $QL$ by SPU.

Under the assumption that the underlying symmetric key encryption is chosen plaintext secure, and the underlying PKSE is secure against chosen keyword attacks with trapdoor privacy~\cite{GolleSW04}, the above generic construction is secure. 

Since the generic construction PPGRT consists of two parts, namely encrypted data and encrypted search index structure, its security mainly depends on two aspects: one is the security of the underlying symmetric key encryption, and the other one is the security of the PKSE. The chosen plaintext security of the symmetric encryption guarantees the secrecy of the underlying haplotype. The security holding against chosen keyword attacks is used to ensure the query haplotype search index (issued by TI) to be hidden from the view of search server. The trapdoor privacy of the PKSE is to protect the haplotype embedded in the searchable trapdoor from being known by the ``curious'' server. The security of the whole PPGRT construction can be proved by following the universally composable model (introduced in~\cite{Canetti01}) with the above security assumption. %
\subsection{Paillier's Encryption System}
Below we review the Paillier Encryption. Please refer to \cite{Paillier99} for more definition and technical details. We let $n=pq$ be a safe number, $\mathcal{B}_{\alpha} \in \mathbb{Z}_{n^2}^*$ be the set of elements of order $n\alpha$, $\mathcal{B}$ be their disjoint union for $\alpha=1, ..., \lambda$, where $p=2p'+1$, $q=2q'+1$, $p,p',q,q'$ are large primes, Carmichael's function $\lambda(n)=lcm(p-1, q-1)$, and set $\lambda(n)$ to be $\lambda$. Suppose $g$ is a base of $\mathcal{B}$, and set $L(x)=\frac{x-1}{n}$. In Paillier encryption system, we set $(n,g)$ as public parameters, and $\lambda$ as secret key. The encryption algorithms works as 
\begin{displaymath}
Enc(x,r): y = g^x \cdot r^n \ mod\ n^2,  
\end{displaymath}
where $y$ is the ciphertext, $x$ is the message, $r$ is a random value, and $x<n$, $r<n$.
The decryption algorithm runs as 
\begin{displaymath}
Dec(y,\lambda): x = \frac{L(y^{\lambda} \ mod\ n^2)}{L(g^{\lambda} \ mod\ n^2)}\ mod\ n,  
\end{displaymath}
where the ciphertext $y< n^2$. 

The Paillier encryption supports the following homomorphic properties: 
\begin{displaymath}
\forall \ \sigma \in \mathbb{Z}^+,\ Dec(y^\sigma, \lambda) = \sigma x\ mod\ n; \\
\forall \ y_1, y_2 \in \mathbb{Z}_{n^2}, \ Dec(y_1 y_2, \lambda) = x_1 + x_2\ mod\ n. 
\end{displaymath}

\subsection{A Basic Concrete Construction}
Following the generic construction, we now build a concrete system for PPGRT (via the usage of LCS) on top of a single keyword equality variant of the searchable encryption scheme~\cite{WWP07}. In the construction, we revise the LCS algorithm (note our LCS algorithm is revised from the one given in http://introcs.cs.princeton.edu/java/96optimization/LCS.java.) so that it only outputs the shared length between a pair of encrypted haplotypes but not the LCS sequence.  
\begin{itemize}
\item {\bf Setup($1^{k}$):} The CI chooses two target collision resistant hash functions $H: \{0,1\}^l \to \mathbb{Z}_n$, $H_0: \mathbb{G}_1 \to \mathbb{Z}_n$, $\sigma \in_R \mathbb{Z}^+$, a cyclic group $\mathbb{G}_1=\left< P \right>$ with order $n$ ($P$ is a base of the group, the computation is based on the modulus $n$), sets $\beta=\sigma \lambda \ mod \ \phi(n^2)$, $\gamma = \sigma L(g^{\lambda} \ mod \ n^2)\ mod\ n$, where $\lambda, g, L$ are parameters in Paillier encryption (please refer to the previous section and~\cite{Paillier99} for more details), and Euler's Totient function $\phi (n) = (p-1)(q-1)$, and $\phi(n^2)=n\phi(n)$. The CI sets system public key as $pk=(1^k, g, H, P, n, \beta, L)$, sets its secret key as $sk=(\sigma, \gamma, \lambda)$, and sends SPU a public key tuple $spk=(1^k, \beta, L, n)$.

\item {\bf GenEDB($sk$, $pk$, $GDB$):} Before delivering a $GDB$ to an SPU, the CI encrypts it as follows. Below we use $Y_i$ to denote a plain haplotype. 
\begin{enumerate}
\item For $i=1$ to ${d}$, the CI runs as 

\begin{enumerate}
\item Given $pk$ and a haplotype $Y_i$ with length $t$, the CI works as follows. 
To randomize each letter, for each $W_j\in \{A,G,C,T,*\}$ ($j\in [1,t]$), the CI randomly chooses new random $\rho_j, \eta_j \in_R \mathbb{Z}_n$, and sets $b_{j}=\rho_j H_0({\eta_j}P)$ and $s_{j}=\rho_j \gamma H(W_j) H_0({\eta_j}P)$. The haplotype's encrypted index sequence $ind_i$ now is represented by a set of $T_i$ in which $T_i^j=(b_j, s_j)$. For convenience, we use a notation $(B, S)$ later, where $|B|=|S|=t$, $B$ stores random factors $b_{j}$ and $S$ is with a set of random $s_{j}$. Note that CI will choose a pair of distinct random factors ($\rho$, $\eta$) for each letter so as to avoid the case where the repeated letters share with the same random pair. For example, for a two-bit string $AA$, CI chooses random factors $\rho_1, \eta_1$ and $\rho_2, \eta_2$ for the first and second $A$, respectively.  

\item The CI also encrypts $Y_i$ by using a symmetric key encryption system $SYE=(SYE.KEY$, $SYE.ENC$, $SYE.DEC)$. It sets $key \gets SYE.KEY(1^k)$, and computes $Z_i=SYE.ENC(Y_i, key)$. We suppose the length of $key$ is identical to that of $Y_i$, such that the symmetric encryption is a perfect one-time pad to $Y_i$. 
\end{enumerate}  

\item The CI finally outputs an $EDB = (ind_i, Z_i)_{i=1}^d$. 
\end{enumerate}   

\item {\bf GenQuery($pk$, $X$):} Before sending its haplotype sequence $X$ to the SPU for relatedness test, the TI encrypts the sequence as follows. Suppose $|X|=m$, and $X=(W_1,...,W_m)$. The TI sets $c_j = g^{H(W_j)}\cdot r_j ^{n} \ mod\ n^2$, and interprets the sequence as $C=(c_1,...,c_m)$, where $r_j \in_R \mathbb{Z}_n$, $c_j$ is a Paillier encryption for $H(W_j)$, and $j\in \{1,...,m\}$. Each $W_j$ is hidden (with uniformly random value $r_j$) by the Paillier probabilistic encryption.   

\item {\bf Test($spk$, $EDB$, $C$):} Given an encrypted $EDB$ and an encrypted haplotype sequence $C$, the SPU first extracts a tuple $(B_z, S_z)$ of an encrypted haplotype $Z_z$ from the $EDB$, where $z \in [1, d]$ and $|S_z|=t$. From $z=1$ to $d$, the SPU runs the algorithm $PPLCS(B_z, S_z, C)$ (see the privacy-preserving longest common subsequence Algorithm~\ref{a1}), and finally outputs the length of the shared positions between $Z_z$ and $C$. Note that we use $B_z[i]$ and $S_z[i]$ to denote the $i$-th element in array $B_z$ and $S_z$, respectively. Note that the LCS algorithm here is revised from the one shared in Princeton Java Library (http://introcs.cs.princeton.edu/java/96optimization/LCS.java.).  

{
\begin{algorithm} 
\caption{Privacy-Preserving Longest Common Subsequence}\label{a1}
\small
\begin{algorithmic}[1]
\Procedure{PPLCS}{$B_z$, $S_z$, $C$}
\BState int $w[0,...,t, 0,...,m]$;

\BState for $i=0$ {to} $t$ {do} $\{$ $w[i,0]=0$; $\}$

\BState for $j=0$ to $m$ {do}  $\{$ $w[0,j]=0$; $\}$

\BState for $i=1$ {to} $t$ {do} $\{$ 
\State  for $j=1$ {to} $m$ {do} $\{$ 

\State 	if (\boxed{L(C[j]^{\beta \cdot B_z[i]}\ mod\ n^2)==S_z[i]\ (mod\ n)}) $\{$ $w[i,j] = 1+ w[i-1, j-1]$; $\}$

\State else if $( w[i-1, j] \ge w[i, j-1] )$ $\{$ $w[i,j]= w[i-1, j]$; $\}$

\State else $\{$ $w[i,j]= w[i, j-1]$; $\}$

\State $\}$

\BState $\}$

\BState return $w[t,m]$; 
\EndProcedure
\end{algorithmic}
\end{algorithm}
}
\end{itemize}
Since the correctness mainly follows that of~\cite{WWP07}, we can have the check below. 

One may verify the equality of $(\ L(C[j]^{\beta \cdot B_z[i]}\ mod\ n^2)==S_z[i]\ (mod\ n)\ )$ to see if the letter $W_i$ in $S_z[i]$ is equal to the letter $W_j$ in the encryption $C[j]$ as in~\cite{WWP07}.  
Since the decryption algorithm of Paillier encryption system can be represented as $\xi (H(x_j)) = \frac{L((C[j]^\xi)^{\lambda}\ mod\ n^2)}{L(g^{\lambda} \ mod\ n^2)} \ (mod\ n)$, we have $\xi (H(x_j)) \cdot L(g^{\lambda} \ mod\ n^2)= L((C[j]^\xi)^{\lambda}\ mod\ n^2)\ (mod\ n)$. Set $\xi=\rho_i \sigma H_0(\eta_i P)$, we have 
\begin{displaymath}
\begin{split}
&\rho_i \sigma  H_0(\eta_i P) (H(x_j))  L(g^{\lambda} \ mod\ n^2) = L((C[j]^{\rho_i  \sigma  H_0(\eta_i P)})^{\lambda}\ mod\ n^2)\ (mod\ n)\\
&\rho_i  H(x_j) H_0(\eta_i P) \sigma  L(g^{\lambda} \ mod\ n^2)  =  L(C[j]^{\rho_i \sigma  H_0(\eta_i P) \lambda}\ mod\ n^2)\ (mod\ n)\\
&\rho_i  H(x_j) H_0(\eta_i P) \gamma  =  L(C[j]^{B[i] \beta}\ mod\ n^2)\ (mod\ n)%
\end{split}
\end{displaymath}
If the letter $x_j$ (on the left hand side of the equation) is equal to the letter $W_j$ embedded in the encryption $C[j]$, we then definitely have $\rho_i  H(x_j) H_0(\eta_i P) \gamma  =  L(C[j]^{B[i] \beta}\ mod\ n^2)\ (mod\ n)$, so that $S[i]  =  L(C[j]^{B[i] \beta}\ mod\ n^2)$ $(mod\ n)$. 
We note that the equation can be checked by SPU without knowing $sk$. 

The concrete construction above is secure assuming the underlying symmetric encryption is secure, the Paillier encryption is secure and \cite{WWP07} is secure with trapdoor privacy in the indistinguishability of ciphertext from random game. We note that the details of the proof follow that of~\cite{WWP07}. 

\subsection{Improvements and Extensions} 
In the previous section, a basic concrete construction for our generic PPGRT is built based on the searchable encryption scheme~\cite{WWP07}. It is not fully secure yet because of suffering from deterministic identifier and offline keyword guessing attacks. Below we present some solutions to tackle the attacks and meanwhile, we show that the construction can be extended to friendly support edit and hamming distance algorithms. 

\subsubsection{Eliminating Deterministic Identifier for H(W)} 
From the construction of the haplotype encrypted index sequence, we can see that $b_j$ and $s_j$ are constructed as $\rho_j H_0({\eta_j}P)$ and $\rho_j \gamma H(W_j) H_0({\eta_j}P)$, respectively. Recall that the tuple ($b_j$, $s_j$) is given to the SPU. Here, a malicious SPU can easily obtain $\delta_j=s_j/b_j=\gamma H(W_j)$. Taking two distinct $\delta_j$ and $\delta_j'$, the malicious SPU can compute $gcd(\delta_j, \delta_j')$ to recover $\gamma$. With knowledge of $\gamma$, it may correctly guess the haplotype sequence, since the message space ($m\in \{A,G,C,T,*\}$) is relatively small in our context. To prevent the attack, the CI may choose to add random factor into the hash function. It can choose a random value, say $s\in \mathbb{Z}_n$, so that $s_j=\rho_j \gamma H'(W_j,s) H_0({\eta_j}P)$, where $H'$ is a new target collision resistance hash function so that $H': \{0,1\}^l \times \mathbb{Z}_n \to \mathbb{Z}_n$. Now, even being able to achieve $\gamma H(W_j, s)$, the malicious SPU may not easily guess what is the input of the hash function. This naive solution, however, incurs another deterministic issue. While the input of the hash function is identical - meaning that a SNPs letter is repeated, the hash output leads to the same value, for example, $H(W_i=A, s)=H(W_j=A, s)$. In such a case, the malicious SPU may make use of some statistical analysis (e.g., 70\% of the repetition occurs for the letter $A$ or $C$) to reveal the whole haplotype sequence. A better countermeasure is to bring more random factors into the construction of $s_j$. Specifically, the CI may set $s_j=(\rho_j \gamma_0  H_0({\eta_j}P)+\gamma_j z_j)H'(W_j,s_j)$, $k_j=a_j z_j$, where $a_j, z_j, s_j\in_R \mathbb{Z}_n$, $\gamma_0=\sigma L(g^{\lambda}\ mod\ n^2) mod\ n$ and $\gamma_j =(a_j L(g^{\lambda}\ mod\ n^2) mod\ n) / \lambda$. Note the $ind_i$ now includes ($B$, $S$, $K$), where $K$ is the set of all $k_j$. Accordingly, the equality check needs to be revised as $(\ L(C[j]^{\beta \cdot B_z[i]+ K_z[i]}\ mod\ n^2)==S_z[i]\ (mod\ n)\ )$. One may check the correctness by setting $\xi = \rho_j \sigma  H_0({\eta_j}P)+a_j z_j/\lambda$. Note we will talk about the random factor $s_j$ later. 

Since the message space of haplotype is relative small, $\{A,G,C,T,*\}$, it is extremely important for CI to produce randomly indistinguishable index sequence in the privacy point of view. Recall that this index sequence can be exactly seen as a searchable trapdoor (in our generic PPGRT construction). Therefore, the privacy of the index sequence is now reduced to the trapdoor privacy. There have been some research works that introduce effective methods to tackle the trapdoor privacy issue, e.g.~\cite{ArriagaTR14}. The crucial idea behind trapdoor privacy is to disable SPU's ability of telling/identifying the relationship between (any) two given trapdoors (with respective keywords inside). It is much like the anonymity feature in the context of identity-based encryption. The premise of doing so is to ensure a fresh randomness to each trapdoor. We here recommend readers to take a public key-based searchable encryption system with trapdoor privacy as an input building block to our generic PPGRT ``compiler'', while attempting to construct a concrete scheme, so as to guarantee the privacy of haplotype encrypted index sequence.

{
\begin{algorithm}[b] 
\caption{Privacy-Preserving Hamming Distance Algorithm}\label{a3}
\small
\begin{algorithmic}[1]
\Procedure{PPLCS}{$B_z$, $S_z$, $C$}
\BState int $seg=0$;

\BState if $(t$ $!=$ $m)$ $\{$ \emph{Please input equal length segments.} $\}$ //Recall that $|C|=m$ which is the test issuer's encrypted sequence, while $|S_z|=t$ which is the genomic index sequence, and $B_z$ is the random factor sequence.

\BState else for $i=1$ {to} $t$ {do} $\{$

\State 	if (\boxed{L(C[i]^{\beta \cdot B_z[i]}\ mod\ n^2)==S_z[i]\ (mod\ n)}) $\{$ $seg++$; $\}$

\BState $\}$

\BState return $seg$; 
\EndProcedure
\end{algorithmic}
\end{algorithm}
}

\subsubsection{Protecting against Offline Keyword Dictionary Attack} 
Since the Paillier encryption algorithm is publicly known, the SPU may launch offline keyword dictionary attack to guess the information of the haplotypes stored in its database. Specifically, the SPU can randomly choose a haplotype sequence $X'=(W_1',...,W_m')$, and further runs the algorithm $GenQuery$ to generate an encrypted sequence $C'$. By running the algorithm $Test$ with input $C'$ and a client's haplotype-hidden index $ind_i$ stored in $EDB$, the SPU may learn how similarity the two sequences share with. Accordingly, the client's haplotype information definitely suffers from a leakage risk. To thwart this possible attack, our system allows the CI to share a secret information $s$, being seen as an extra input for the hash function, with the TI via a well-studied and efficient Diffie-Hellman key exchange protocol (in DES way) with EBC mode, so that the TI will construct each encrypted element $c_j$ (of the corresponding letter $W_j$) as $H(W_j, s)$. Accordingly, the secret information $s$ has to be put in the index element $s_j$ by the CI as well. The SPU will fail to launch effective offline keyword dictionary attack, since it cannot compute a valid hash value without knowledge of the shared secret $s$. The key exchange protocol above can be seen as a permission granted from the CI, so that the TI can proceed to the test phase. Imagine that in practice, a public tester usually needs to request a test permission from the EDB owner, i.e. the CI, even the EDB is outsourced to the SPU. If the CI grants the permission, it will share the secret information with the TI; otherwise, nothing will be shared.  

To keep consistency with the revision introduced in the previous subsection, a random factor $s$ is replaced with a $\hat{s}=(s_1,....,s_t)$, where $s_j\in_R \mathbb{Z}_n$ and $j \in [1,t]$. We note that there is no need for the CI and the TI to share with the whole vector $\hat{s}$ in the exchange protocol. For example, the CI may leverage the equation $a x_j+b$ to calculate the value $s_j$, where $a, b\in_R \mathbb{Z}_n$ and $x_j$ is some ``common'' information that the TI knows as well (e.g. $x_j=H(p_j, W_j)$, where $p_j$ is the current position of the letter $W_j$ on the sequence). In this case, the CI will only need to share $a,b$ with the TI in the key exchange protocol.

{
\begin{algorithm}[t]
\caption{Privacy-Preserving Edit Distance Algorithm}\label{a2}
\small
\begin{algorithmic}[1]
\Procedure{PPLCS}{$B_z$, $S_z$, $C$}
\BState int $len1 = m$; int $len2 = t$; int $len=0$;

\BState if ($len1 < len2$) $\{$ $len=len2$; $\}$

\BState else $\{$ $len=len1$; $\}$

\BState int [][] $dp$ = new int [$len1+1$] [$len2+1$];

\BState for $i=0$ {to} $len1$ {do} $\{$ $dp[i][0]=i$; $\}$

\BState for $j=0$ {to} $len2$ {do} $\{$ $dp[0][j]=j$; $\}$

\BState for $i=0$ {to} $len1-1$ {do}
  
\State  for $j=0$ {to} $len2-1$ {do} $\{$ 

\State 	if (\boxed{L(C[i]^{\beta \cdot B_z[j]}\ mod\ n^2)==S_z[j]\ (mod\ n)}) $\{$ $dp[i+1][j+1]=dp[i][j]$; $\}$

\State else $\{$ 

\State int $replace = dp[i][j]+1$; 

\State int $insert = dp[i][j+1]+1$;

\State int $delete = dp[i+1][j]+1$;

\State int $min = replace > insert ? insert : replace$;

\State $min = delete > min? min : delete$;

\State $dp[i+1][j+1] = min$; 

\State $\}$

\BState $\}$

\BState return $len-dp[len1][len2]$; 
\EndProcedure
\end{algorithmic}
\end{algorithm}
}

\subsubsection{Hamming Distance based PPGRT} 
We state that our concrete construction is naturally compatible with edit distance and hamming distance algorithms. The intuition is to see the equation highlighted in the box in Algorithm~\ref{a1} as a condition for distance counting -- that is, if the equation does not hold (indicating a mismatch), we proceed to a distance adding operation -- and, meanwhile, the final output is set to only show the shared length of two given strings. The privacy-preserving hamming distance algorithm is shown in~Algorithm~\ref{a3}, in which we assume that each segment of the test issuer's encrypted sequence (being taken into the algorithm) is with the same length of the genomic index. 

\subsubsection{Privacy Preserving via Edit Distance} 
The privacy-preserving variant for edit distance algorithm is presented in~Algorithm~\ref{a2}. We note that the output of our privacy-preserving edit distance algorithm is somewhat different from that of the original edit distance algorithm -- outputting the number of the ``difference'' of two strings.

\begin{table*}[b]
\centering
\footnotesize
\begin{tabular}{|c|c|}
\hline
 Item & Description \\
\hline
Computer & MacBook Pro OS X Yosemite 10.10.5 \\
\hline
CPU & 2.70GHz Intel(R) Core(TM) i5-5257U \\
\hline
Memory & DDR 3, 8 GB, 1867 MHz \\
\hline
Flash Storage & 120.11 GB \\
\hline
Programming Platform& Eclipse IDE Mars.2 Release (4.5.2) \\
\hline
Programming Language& Java Version 1.8.0\_60 \\
\hline
\end{tabular}
\vspace{-0.2cm}
\caption{Test Bed Details}
\label{tb1}
\end{table*}

\section{Performance Evaluation}
\label{sec:evaluation}
We also perform an efficiency evaluation for our protocols, aiming to demonstrate that the overhead imposed to both CI and TI are appreciably low. We implement the concrete construction using MD5 as the cryptographic hash function, a 1024-bit Diffie-Hellman key agreement (for sharing a secret hash function input between CI and TI only), and Paillier cryptosystem with 2048-bit moduli. The experimental test bed is shown in Table~\ref{tb1} below. We here use real SNPs database to test the running time and accuracy of our privacy-preserving algorithms. We show the SNPs sample details in Table~\ref{tb2}.
\begin{table*}[t]
\centering
\footnotesize
\begin{tabular}{|c|c|}
\hline
 Item & Description \\
\hline
Database & 1000 Genomes\footnote{\url{http://www.1000genomes.org/}} \\
\hline
Phase & 1000 Genomes phase 3 release \\
\hline
Dataset Name & NA12828 \\
\hline
Biosample ID & SAME125076 \\
\hline
Superpopulation & European (EUR) \\
\hline
Population Code& CEU \\
\hline
Population Description & Utah residents (CEPH) with Northern and Western European ancestry \\
\hline
Gender & Female\\
\hline
\end{tabular}
\caption{Experimental Sample. $^*$The size of dataset is total 565.8 MB, including 4,743,960 genomic strings. But we only choose to use the first 10,000 strings of that, with the size about 1.1 MB, for the our test.}\label{tb2}
\end{table*}

We further assume that there is a pair of haplotypes, one encrypted by CI, and the other encrypted by TI: both encryptions will be sent to SPU for two groups of test - one group of test is for original (non-privacy-preserving) LCS, edit and hamming distance algorithms, and the other is for privacy-preserving ones. Note that both test groups will intake the identical haplotype sample. Our pair of haplotypes are in the ``A, G, C, T'' format with the length 36$\times$500 letter bits (note in this paper we refer one letter to as a letter bit, e.g. ``AG'' are two letter bits). For matching 1000 Genomes' data format, we cut a total 18,000 letter bits haplotype into 500 segments of which contains 36 letter bits. Note that the running times of each algorithm are averaged over 100 executions. 

\begin{figure*}[t]
\centering
\includegraphics[width=4in]{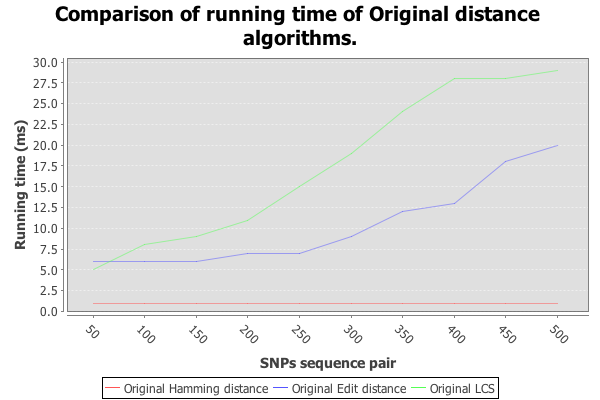}
\caption{Running Time of Non-Privacy-Preserving Distance Algorithms}
\label{fg2}
\end{figure*}

\begin{figure*}
    \centering
    \includegraphics[width=4.in]{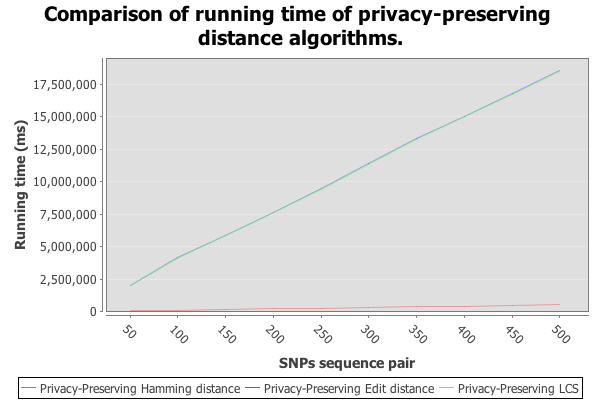} 
    \caption{Running Time of Storage and Processing Unit} 
     \label{fg3} 
\end{figure*}

The running time of non-privacy-preserving distance algorithms is depicted in Figure~\ref{fg2}. It can be seen that hamming distance algorithm significantly outperforms edit distance algorithm, while the LCS suffers from the worst performance. The running time test here appropriately show the efficiency of the three algorithms.   
The running time of privacy-preserving hamming distance algorithm (in Figure ~\ref{fg3}) only takes approximately 900s to finish the test by intaking two encrypted haplotypes with 36$\times$500 SNPs letter bits, while the privacy-preserving LCS and edit distance (nearly overlapped) requires over $10^4$s. Accordingly, dealing with phased haplotype, it is better to leverage hamming distance algorithm as a secure building block for our PPGRT.   

We here state that the matching result of using privacy-preserving algorithms is identical to that of using original algorithms by inputing the same sample, 36$\times$500 SNPs. Note that please refer to our supplement test materials. We therefore conclude that the privacy-preserving algorithms maintain the exact test accuracy compared to the ones without any privacy protection.

However, the privacy-preserving algorithms, specially for edit distance and LCS, suffer from huge efficiency loss as a price to maintain privacy. This research work creates an interesting open problem on how to improve the test efficiency without loss of privacy.    

\section{Conclusion}
\label{sec:conclusions}
This research work introduced an interesting observation about how to generically construct a privacy preserving relatedness test protocol based on public key searchable encryption, and next proposed a concrete construction as well as its performance evaluation. The evaluation showed that our construction built on top of hamming distance algorithm is very efficient, while the designs for edit distance and LCS suffer from efficiency loss. The simulation also leverages real-world genetic database (1000 Genomes) to show the test accuracy of the system. The efficiency improvement of the construction will be seen as a future work. 

This paper also leave the academic and industrial communities followings interesting open problems - to improve the efficiency, can we make use of a symmetric searchable encryption to achieve the same GRT test function without loss of privacy; is it possible to directly use a normal but not reversed searchable encryption technique in our context; how can we protect the haplotype (encrypted and outsourced to SPU) even in the case where SPU colludes with a group of TIs; how does CI ``update'' EDB (which is outsourced to SPU) but also ``control'' which parts of the EDB to be ``performed'' test tasks by the SPU; what if there are different encryption formats in the back-end of the SPU, how does it perform efficient and effective test.  

\descr{Acknowledgments.} The authors wish to thank Alexandros Mittos for presenting the paper at the GenoPri workshop. This work is supported by a Google Faculty Award grant and EU Project H2020-MSCA-ITN ``Privacy \& Us'' (Grant No. 675730).

\small
\bibliographystyle{abbrv}

\end{document}